\documentclass[aps,prc,groupedaddress,showpacs,preprintnumbers,twocolumn]{revtex4}
\usepackage{
color}
\usepackage{epsfig}
\usepackage{longtable}
\begin{document}

\title{Low-lying $\Omega$ states with negative parity in an extended quark model with Nambu-Jona-Lasinio
interaction}

\author{C. S. An$^{1}$}\email{ancs@ihep.ac.cn}

\author{B. S. Zou$^{2,1}$}\email{zoubs@ihep.ac.cn}

\affiliation{1.Institute of High Energy
Physics and Theoretical Physics Center for Science Facilities,
Chinese Academy of Sciences, Beijing 100049, China\\
2.  State Key Laboratory of Theoretical Physics, Kavli Institute for
Theoretical Physics China, Institute of Theoretical Physics, Chinese
Academy of Sciences, Beijing 100190, China }

\thispagestyle{empty}

\date{\today}

\begin{abstract}

Here we investigate mixing of the low-lying three- and five-quark
$\Omega$ states with spin-parity quantum numbers $\frac{1}{2}^{-}$
and $\frac{3}{2}^{-}$, employing the quark-antiquark creation
triggered by Nambu-Jona-Lasinio (NJL) interaction. Wave functions of
the three- and five-quark configurations are constructed using the
extended constituent quark model, within which the hyperfine
interaction between quarks is also taken to be the NJL interaction
induced one. Numerical results show that the NJL-induced pair
creation results in vanishing mixing between three- and five-quark
$\Omega$ configurations with spin-parity $1/2^{-}$, but mixing
between three- and five-quark $3/2^{-}$ $\Omega$ states should be
very strong. And the mixing decreases energy of the lowest $3/2^{-}$
$\Omega$ state to be $1785\pm25$~MeV, which is lower than energy of
the lowest $1/2^{-}$ state in this model. This is consistent with
our previous predictions within the instanton-induced
quark-antiquark creation model.

\end{abstract}


\maketitle

\section{Introduction}
\label{sec:intro}

Recently, spectrum of low-lying $\Omega$ resonances with negative
parity was investigated employing an extended constituent quark
model~\cite{s^3,An:2013zoa}, within which the $\Omega$ resonances
were considered as admixtures of three- and five-quark components,
and the hyperfine interaction between quarks was taken to be of
three different kinds, namely, one gluon exchange
(OGE)~\cite{Isgur:1979be,Chao:1980em,Isgur:1977ef,Capstick:2000qj},
Goldstone boson exchange (GBE)~\cite{Glozman:1995fu}, and
instanton-induced interaction
(INS)~\cite{Loring:2001kx,Blask:1990ez,Klempt:1995ku,Koll:2000ke}.
In Ref.~\cite{An:2013zoa}, mixing of three- and five-quark $\Omega$
states was calculated by treating the $q\bar{q}$ creation mechanism
as the one induced by instanton interaction. It is shown that the
mixing between three- and five-quark components in $\Omega$
resonances with spin-parity $1/2^{-}$ is very small and negligible,
but in the $3/2^{-}$ $\Omega$ resonances the mixing is very strong,
and the mixing decreases the energy of the lowest $3/2^{-}$ state to
be around $1750\pm50$ MeV. It is very interesting that this energy
is lower than energy of the lowest spin-parity $1/2^{-}$ $\Omega$
resonance.

As shown in~\cite{An:2013zoa}, the instanton quark-antiquark pair
creation precludes transitions between $s^{3}$ and $s^{4}\bar{s}$
configurations, while the instanton-induced hyperfine interaction
between a quark and an antiquark could lead to mixing between
five-quark $\Omega$ configurations with light quark-antiquark pair
and $s\bar{s}$ pair~\cite{s^3}. Therefore, Once we take the
instanton-induced hyperfine interaction and quark-antiquark pair
creation simultaneously, mixing between between $s^{3}$ and
$s^{4}\bar{s}$ configurations will not vanish. But if the hyperfine
interaction between quarks is chosen as OGE or GBE, the
instanton-induced quark-antiquark pair creation mechanism cannot
result in mixing between $s^{3}$ and $s^{4}\bar{s}$ configurations.
Generally, even if probability of $s^{4}\bar{s}$ component in
$\Omega$ resonances may be small, but it should not be exactly $0$.
This may indicate that once we take the instanton-induced
quark-antiquark pair creation, we have to keep the hyperfine
interaction between quarks being based on the same model.

In the present work, we try to calculate mixing between three- and
five-quark components in low-lying $\Omega$ resonances with negative
parity using the Nambu-Jona-Lasinio (NJL) approach~\cite{NJL1,NJL2},
which was originally constructed for nucleons that interact via an
effective two-body contact interaction, and later developed to
include the quark freedom~\cite{Eguchi:1976iz}. Analogous to the
instanton
interaction~\cite{'tHooft:1976fv,Shifman:1979uw,Schafer:1996wv}, the
NJL model can describe various aspects of QCD related to the
dynamical and explicit breaking of chiral symmetry and the axial
anomaly very well~\cite{Hatsuda:1994pi}. As discussed above, here we
take the hyperfine interactions between quarks to be also the NJL
induced one for model consistency.

The present paper is organized as follows. In Section \ref{sec:frame}, we
present our theoretical framework, which includes explicit forms of the
NJL-induced quark-quark hyperfine interactions, and
quark-antiquark pair creation mechanism.  Numerical results
for the spectrum of the states under study and the mixing of three- and
five-quark configurations in our model are shown in Section
\ref{sec:result}. Finally, Section \ref{sec:end} contains a brief conclusion.

\section{Theoretical Framework}
\label{sec:frame}

In the present model, the Hamiltonian is almost all the same as that using in~\cite{s^3,An:2013zoa}
only except for the parts describe the quark-quark hyperfine interaction and
mechanism for transition between three- and five-quark components
in $\Omega$ resonances. For completeness, here we also repeat the same parts. The Hamiltonian
describing $\Omega$ resonances as admixtures of three- and five-quark components is of
the following form:
\begin{equation}
  H=
  \pmatrix{ H_{3}  &  T_{\Omega_{3}\leftrightarrow\Omega_{5}}  \cr
    T_{\Omega_{3}\leftrightarrow\Omega_{5}}      &  H_{5}}
  \,,
  \label{ham}
\end{equation}
where $H_{3}$ and $H_{5}$ are the Hamiltonian for three-quark and five-quark systems,
respectively, and $T_{\Omega_{3}\leftrightarrow\Omega_{5}}$ denotes the
transition between three- and five-quark systems. Here we discuss the diagonal and
non-diagonal terms of the Hamiltonian (\ref{ham}) in Secs.~\ref{dia} and ~\ref{ndia},
respectively.

\subsection{Diagonal terms of the Hamiltonian}
\label{dia}

The Hamiltonian for a $N$-particle system in the constituent quark model
can be written as the following form
 \begin{eqnarray}
   H_{N}&=&H_{o}+H_{\textit{\footnotesize hyp}}+\sum_{i=1}^{N}m_{i}\,,
\label{hn}
\end{eqnarray}
where $H_{o}$ and $H_{\textit{\footnotesize hyp}}$ represent the Hamiltonians
for the quark orbital motion and for the hyperfine interactions between
quarks, respectively, $m_{i}$ denotes the constituent mass of the $ith$
quark. The first term $H_{o}$ can be written as a sum of the kinetic energy term
and the quark confinement potential as
\begin{equation}
  H_{o}=\sum_{i=1}^N {\vec{p}_i^2\over 2 m_{i}}+\sum_{i<j}^N
  V_{\textit{\footnotesize conf}}(r_{ij})\,.
  \label{ho}
\end{equation}
In~\cite{s^3} the quark confinement potential was taken to be
\begin{equation}
  V_{conf}(r_{ij})=-\frac{3}{8}\lambda_i^C\cdot\lambda_j^C
  \left[C^{(N)}(\vec{r}_i-\vec{r}_j)^2+V_0^{(N)}\right]\,,\label{conf}
\end{equation}
where $C^{(N)}$ and $V_{0}^{(N)}$ are constants. In principle these two
constants can differ for three- and five-quark configurations. $H_{hyp}$
denotes the hyperfine interaction between quarks, here we take $H_{hyp}$
to be the NJL interaction induced one. The NJL interaction between quarks
can be described by
\begin{equation}
 \mathcal{L}_{NJL}=\frac{1}{2}g_{s}\sum_{a=0}^{8}\big[(\bar{q}\lambda^{a}q)^{2}+(\bar{q}i\lambda^{a}\gamma_{5}q)^{2}\big]
 \label{njl}
\end{equation}
where $\lambda^{a}$~($a=1\ldots8$) are the Gell-Mann matrices in the flavor $SU(3)$ space,
and $\lambda^{0}=\sqrt{\frac{2}{3}}\mathcal{I}$, with $\mathcal{I}$ the unit matrix in the
three dimensional flavor space.
In the nonrelativistic approximation, the NJL induced quark-quark interaction can be obtained
as
\begin{equation}
 H_{qq}^{NJL}=\sum_{i<j}^{N}\sum_{a=0}^{8}\hat{g}_{ij}\lambda_{i}^{a}\lambda_{j}^{a}[1+\frac{1}{4m_{i}m_{j}}\hat{\sigma}_{i}
 \cdot(\vec{p}^{\prime}_{i}-\vec{p}_{i})\hat{\sigma}_{j}\cdot(\vec{p}^{\prime}_{j}-\vec{p}_{j})]\,,
 \label{insqq}
\end{equation}
where $\hat{g}_{ij}$ is an operator which distinguishes the coupling strength between two light
quarks $g_{qq}$, one light and one strange quarks $g_{qs}$ and two strange quarks $g_{ss}$.
In principle, once the $SU(3)$ breaking effects are taken into account, the three coupling strength
should be different. One may find that the present hyperfine interaction is very similar
to the one mediated by Goldstone boson exchange, which includes the pseudoscalar and scalar mesons exchange~\cite{Glozman:1995fu}.
Therefore, here we take the relationship of the three different coupling strength as in~\cite{Glozman:1995fu}
\begin{equation}
 g_{qq}\colon g_{qs}\colon g_{ss}=1\colon\frac{m}{m_{s}}\colon\frac{m^{2}}{m_{s}^{2}}\,,
\end{equation}
where $m$ and $m_{s}$ represent the constituent masses of the light and strange quarks.
The empirical value for the constituent mass of light quark is in the range $300\pm50$~MeV,
and that for strange quark is $\sim120-200$~MeV higher than $m$. In the traditional
$qqq$ constituent quark model, $m$ is often taken to be $320-340$~MeV~\cite{Glozman:1995fu}.
In the present case, since we take the baryons to be admixtures of three- and five-quark
components, in general, the constituent quark mass should be lower than that in the three-quark model.
Accordingly, here we take the value $m=310$~MeV for the light quarks, and $m_{s}=460$~MeV for
the strange quark.

Generally, one can divide the interaction~(\ref{insqq}) by different spin dependences~\cite{pn}. If we
neglect the tensor term in the quark-quark interaction, the NJL-induced hyperfine interaction between quarks $H_{hyp}^{NJL}$
should be the following form
\begin{equation}
H_{hyp}^{NJL}=\sum_{i<j}^{N}\sum_{a=0}^{8}\hat{g}_{ij}\lambda_{i}^{a}\lambda_{j}^{a}(1-\frac{1}{12}\hat{\sigma}_{i}
 \cdot\hat{\sigma}_{j})\,,
\end{equation}
Accordingly, one can obtain the three coupling strength by reproducing the mass splitting between
$\Sigma$ and $\Lambda$ baryons
\begin{equation}
 g_{qq}=69~MeV,~g_{qs}=51~MeV,~g_{ss}=38~MeV\,.
\end{equation}

As discussed in~\cite{An:2013zoa}, there are two low-lying $\Omega$ resonances with negative
parity in $N\leq2$ band within the $qqq$ three-quark model~\cite{Glozman:1995fu,Chao:1980em,Pervin:2007wa},
one is with spin $1/2$, and the other $3/2$, correspond to the two first orbitally excited
states of $\Omega(1672)$.
And these two states
should be degenerate in a given hyperfine interaction model if the L-S coupling hyperfine interaction
is not taken into account.
The matrix elements of the sub-matrix $H_{3}$ in~(\ref{ham}) obtained by using the OGE, GBE and INS hyperfine
interactions in~\cite{An:2013zoa} are~$\langle H_{3}^{OGE}\rangle_{\frac{1}{2}^{-}}=\langle
H_{3}^{OGE}\rangle_{\frac{3}{2}^{-}}=2020$~MeV,~$\langle
H_{3}^{GBE}\rangle_{\frac{1}{2}^{-}}=\langle
H_{3}^{GBE}\rangle_{\frac{3}{2}^{-}}=1991$~MeV, and
~$\langle
H_{3}^{INS}\rangle_{\frac{1}{2}^{-}}=\langle
H_{3}^{INS}\rangle_{\frac{3}{2}^{-}}=1887$~MeV, respectively.
In present work, with the above given NJL-induced hyperfine interaction
strength, by reproducing the
mass of the ground state $\Omega(1672)$, one can obtain that $V_{0}^{(3)}=-188$~MeV,
which is smaller than the value $-140$~MeV in GBE model. With these parameters,
we obtain that the matrix elements of $H_{3}$ in present model are
~$\langle
H_{3}^{NJL}\rangle_{\frac{1}{2}^{-}}=\langle
H_{3}^{NJL}\rangle_{\frac{3}{2}^{-}}=1942$~MeV.

On the other hand, to get the value of the parameter for the
$V_{0}^{(5)}$, we fit the lowest five-quark $\Omega$ configuration
to be $\sim1810$~MeV, which value was proposed to be the energy of
the lowest $K\Xi$ bound state with spin-parity $1/2^{-}$~\cite{Wangwl}, this
method yields $V_{0}^{(5)}=-294$~MeV, which is also smaller than the
value $-269$~MeV in GBE model.

\subsection{Non-diagonal terms of Hamiltonian}
\label{ndia}

The non-diagonal term $T_{\Omega_{3}\leftrightarrow\Omega_{5}}$ depends on
the explicit quark-antiquark pair creation mechanism. In present work, we
take the quark-antiquark pair
creation mechanism to be the one based on a nonrelativistic reduction of
the amplitudes found from the NJL interaction. One may find that the two
terms in Eq.~(\ref{njl}) just correspond to the quark-antiquark pairs with
quantum numbers $0^{+}$ and $0^{-}$ creation, respectively. Accordingly, in present case,
only the second term will contribute.
And in the nonrelativistic limit, the second term in Eq.~(\ref{njl}) reduces to
\begin{equation}
 \hat{T}_{q\bar{q}}=-\frac{2}{3m_{s}}g_{qs}\xi_{f}^{\dag}\hat{\sigma}\cdot(\vec{p}_{i}-\vec{p}_{f})\xi_{i}\xi_{q}\mathcal{I}\eta_{\bar{q}}
\label{tqq}
 \end{equation}
for light $q\bar{q}$ creation and
\begin{equation}
 \hat{T}_{s\bar{s}}=\frac{1}{3m_{s}}g_{ss}\xi_{f}^{\dag}\hat{\sigma}\cdot(\vec{p}_{i}-\vec{p}_{f})\xi_{i}\xi_{s}\mathcal{I}\eta_{\bar{s}}
\label{tss}
 \end{equation}
for $s\bar{s}$ creation, where $\xi_{f(i)}$ and $\vec{p}_{f(i)}$ denotes the final (initial) spin and
momentum operators of quark which emits a $q\bar{q}$ or $s\bar{s}$ pair, $\xi_{q(s)}$ is the spin operator
of the created light (strange) quark and $\eta_{\bar{q}(\bar{s})}$ the spin operator of the created light (strange)
antiquark.

 If we treat the other two quarks as spectators, then the
non-diagonal term $T_{\Omega_{3}\leftrightarrow\Omega_{5}}$ in Hamiltonian~(\ref{ham}) can be obtained as
\begin{eqnarray}
 T_{\Omega_{3}\leftrightarrow\Omega_{5}}^{q\bar{q}}&=&-
 \frac{2}{3m_{s}}g_{qs}\sum_{i=1}^{3}\sum_{j\neq i}^{4}\mathcal{C_{F}}\mathcal{C_{S}}
 \mathcal{C_{C}}\mathcal{C_{O}}\nonumber\\
 &&\xi_{i}^{\prime\dag}\hat{\sigma}
 \cdot(\vec{p}_{i}-\vec{p}_{i}^{\prime})\xi_{i}\xi_{j}\mathcal{I}\eta_{\bar{q}}\\
 T_{\Omega_{3}\leftrightarrow\Omega_{5}}^{s\bar{s}}&=&
 \frac{1}{3m_{s}}g_{ss}\sum_{i=1}^{3}\sum_{j\neq i}^{4}\mathcal{C_{F}}\mathcal{C_{S}}
 \mathcal{C_{C}}\mathcal{C_{O}}\nonumber\\
 &&\xi_{i}^{\prime\dag}\hat{\sigma}
 \cdot(\vec{p}_{i}-\vec{p}_{i}^{\prime})\xi_{i}\xi_{j}\mathcal{I}\eta_{\bar{s}}
\end{eqnarray}
for transitions $sss\longleftrightarrow sssq\bar{q}$ and $sss\longleftrightarrow ssss\bar{s}$,
respectively. Where $\mathcal{C_{F}}$, $\mathcal{C_{S}}$, $\mathcal{C_{C}}$ and
$\mathcal{C_{O}}$ are operators for the calculation of the corresponding
flavor, spin, color and orbital overlap factors, respectively.

\section{Numerical results}
\label{sec:result}

As what we have done in~\cite{s^3,An:2013zoa}, to show our numerical results
clearly, here we denote the two three-quark configurations as $|3, \frac{1}{2}^{-}\rangle$
and $|3,\frac{3}{2}^{-}\rangle$, respectively,
and five-quark configurations with spin-parity quantum number $1/2^{-}$ as
\begin{eqnarray}
|5,\frac{1}{2}^{-}\rangle_{1}&=&|s^{3}q([4]_{X}[211]_{C}[31]_{FS}
[31]_{F}[22]_{S})\otimes \bar{q}\rangle\,,\nonumber\\
|5,\frac{1}{2}^{-}\rangle_{2}&=&|s^{3}q([4]_{X}[211]_{C}[31]_{FS}
[31]_{F}[31]_{S})\otimes \bar{q}\rangle\,,\nonumber\\
|5,\frac{1}{2}^{-}\rangle_{3}&=&|s^{3}q([4]_{X}[211]_{C}[31]_{FS}
[4]_{F}[31]_{S})\otimes \bar{q}\rangle\,,\nonumber\\
|5,\frac{1}{2}^{-}\rangle_{4}&=&|s^{4}([4]_{X}[211]_{C}[31]_{FS}
[4]_{F}[31]_{S})\otimes \bar{s}\rangle\,,
\label{num12}
\end{eqnarray}
and those with spin-parity quantum number $3/2^{-}$ as
\begin{eqnarray}
|5,\frac{3}{2}^{-}\rangle_{1}&=&|s^{3}q([4]_{X}[211]_{C}[31]_{FS}
[31]_{F}[31]_{S})\otimes \bar{q}\rangle\,,\nonumber\\
|5,\frac{3}{2}^{-}\rangle_{2}&=&|s^{3}q([4]_{X}[211]_{C}[31]_{FS}
[31]_{F}[4]_{S})\otimes \bar{q}\rangle\,,\nonumber\\
|5,\frac{3}{2}^{-}\rangle_{3}&=&|s^{3}q([4]_{X}[211]_{C}[31]_{FS}
[4]_{F}[31]_{S})\otimes \bar{q}\rangle\,,\nonumber\\
|5,\frac{3}{2}^{-}\rangle_{4}&=&|s^{4}([4]_{X}[211]_{C}[31]_{FS}
[4]_{F}[31]_{S})\otimes \bar{s}\rangle\,.
\label{num32}
\end{eqnarray}

The main parameters involved in the transitions between three- and five-quark
$\Omega$ states are the ratio of harmonic oscillator parameters $R_{35}=\omega_{5}/\omega_{3}$,
and the NJL interaction strength $g_{qs}$ and $g_{ss}$. We present the numerical results
by taking the parameters to be empirical values in Sec.~\ref{fixed}, and those by treating
the the ratio $R_{35}$ and interaction strength $g_{qs}$ and $g_{ss}$ to be free
parameters in Sec.~\ref{free}.

\subsection{Numerical results with fixed parameters}
\label{fixed}

Firstly, we take a tentative value $R_{35}=\sqrt{5/6}$~\cite{An:2013zoa}, and the values for
NJL interaction strength given in Sec.~\ref{dia} to show the mixing between three- and
five-quark $\Omega$ states within the NJL-induced quark-antiquark pair creation model.
With the notations in Eqs.~(\ref{num12}) and~(\ref{num32}), the matrix elements
of the Hamiltonian~(\ref{ham}) including $H_{3}$, $H_{5}$ and $T_{\Omega_{3}\leftrightarrow\Omega_{5}}$ are
{\footnotesize
\begin{eqnarray}
\langle H^{NJL}\rangle_{1/2}=
\pmatrix{  1942.0   &     0       &  0         &  0       &  0      \cr
           0        &     1809.9  &  0    &  0    &  0      \cr
           0        &     0       &   1816.2   & -6.1    &  0      \cr
           0        &     0       &  -6.1     &  2254.8  &  0      \cr
            0       &     0       &   0        &  0       &  2474.7.0 \cr} \,\label{engnjl1} \\
\langle H^{NJL}\rangle_{3/2}=
\pmatrix{  1942.0   &     38.5    &  -60.9    &  -27.2   &  20.3      \cr
           38.5     &     1816.2  &   0    &    -6.1     &  0      \cr
          -60.9    &     0   &   1821.5   & 0    &  0      \cr
          -27.2     &    -6.1      &  0     &  2254.8  &  0      \cr
           20.3       &     0       &   0        &   0      &  2474.7 \cr} \,.\label{engnjl2}
\end{eqnarray}
}
\begin{table}[t]
\caption{\footnotesize Energies and the
corresponding probability amplitudes of three- and five-quark
configurations for the obtained $\Omega$ states in the NJL-induced
hyperfine interaction model. The upper
and lower panels are for states with quantum numbers
$\frac{1}{2}^{-}$ and $\frac{3}{2}^{-}$, respectively,
and for each panel, the first row shows the energies in MeV, others show
the probability amplitudes.
\label{num1}}
\renewcommand
\tabcolsep{0.27cm}
\renewcommand{\arraystretch}{1.6}
\scriptsize
\vspace{0.5cm}
\begin{tabular}{c|ccccc}
\hline\hline
 $\frac{1}{2}^{-}$
&  1810       &   1816      &   1942      &   2255     &   2475
\\
\hline

$|3,\frac{1}{2}^{-}\rangle$
&  0.0000     &   0.0000    &   1.0000    &   0.0000   &  0.0000
\\

$|5,\frac{1}{2}^{-}\rangle_{1}$
&  1.0000     &   0.0000    &  0.0000    &   0.0000   &   0.0000

\\

$|5,\frac{1}{2}^{-}\rangle_{2}$
&  0.0000     &   0.9999    &   0.0000    &   -0.0140   &  0.0000

\\

$|5,\frac{1}{2}^{-}\rangle_{3}$
&  0.0000     &   0.0140    &   0.0000    &   0.9999   &   0.0000

\\

$|5,\frac{1}{2}^{-}\rangle_{4}$
&  0.0000     &   0.0000    &   0.0000    &   0.0000  &   1.0000

\\

\hline
 $\frac{3}{2}^{-}$
&  1786       &   1818      &   1972      &   2257     &   2475

     \\
\hline

$|3,\frac{3}{2}^{-}\rangle$
&  0.4227     &   -0.0354    &  -0.9002    &   0.0905   &   0.0389

\\

$|5,\frac{3}{2}^{-}\rangle_{1}$
& -0.5385     &   -0.8135    &   -0.2185    &  0.0217   &   0.0023

\\

$|5,\frac{3}{2}^{-}\rangle_{2}$
&  0.7286     &  -0.5803    &  0.3635    &  -0.0127   &   -0.0036

\\

$|5,\frac{3}{2}^{-}\rangle_{3}$
&  0.0175     &   -0.0136    &   -0.0916    &   -0.9955    &  -0.0049

\\

$|5,\frac{3}{2}^{-}\rangle_{4}$
& -0.0125          &   0.0011         &   0.0364         &   -0.0085         &  0.9992

\\

\hline
\hline
\end{tabular}
\end{table}

As we can see in Eq.(\ref{engnjl1}), the NJL-induced quark-antiquark
pair creation does not contribute to transitions between three- and
five-quark $\Omega$ states with spin $1/2$. This is because of that
the quark-antiquark pairs in corresponding five-quark $\Omega$
states are with the quantum number $^{3}S_{1}$, while the created
quark-antiquark pair in NJL approach should be with the quantum
number $^{1}S_{0}$, therefore, transitions between the studied spin
$1/2$ three- and five-quark configurations vanish in the NJL-induced
pair creation model. One may find this is consistent with the
results obtained in~\cite{An:2013zoa} using the instanton pair
creation model. In Ref.~\cite{An:2013zoa}, although the obtained
mixing between three- and five-quark configurations with spin 1/2 is
not 0, it is very small and negligible, in fact, the mixing is
proportional to $1/m-1/m_{s}$ , with $m$ and $m_{s}$ the constituent
masses of the light and strange quarks, so the mixing will also be 0
in the flavor $SU(3)$ limit. For the non-diagonal matrix elements
from transitions between three- and five-quark spin $3/2$ $\Omega$
states, Eqs.~(\ref{engnjl2}) shows that the transition matrix
elements are smaller than those caused by instanton-induced
quark-antiquark creation~\cite{An:2013zoa}.

On the other hand, as shown in Eqs.~(\ref{engnjl1}) and~(\ref{engnjl2}),
the NJL-induced hyperfine interaction between quarks leads
to only two very small nonvanishing non-diagonal matrix elements for both spin $1/2$ and
$3/2$ cases. This is the same as the results obtained within GBE hyperfine interaction
model~\cite{s^3}. As we have discussed in Sec.~\ref{sec:frame}, the present hyperfine
interaction model is very similar to the GBE model.

Diagonalization of Eqs.~(\ref{engnjl1}) and~(\ref{engnjl2}) results
in the numerical results of the energies and corresponding
probability amplitudes of three- and five-quark configurations for
the obtained $\Omega$ states shown in Table~\ref{num1}. As shown in
this table, mixing between three- and five-quark spin $3/2$ $\Omega$
configurations are very strong, while since the transition matrix
elements listed in Eq.~(\ref{engnjl2}) are not large, so the mixing
does not decreases energy of the lowest state so much as that
obtained in~\cite{An:2013zoa}. Nevertheless, the obtained energy of
the lowest spin $3/2$ state is lower than energy of the lowest spin
$1/2$ state, this is consistent with our previous results obtained
within the instanton-induced quark-antiquark pair creation
model~\cite{An:2013zoa}. And one can find that the mixing between
three-quark $\Omega$ state and the $s^{4}\bar{s}$ is very small, the
largest mixing appears in the third state with probability of
$s^{4}\bar{s}$ $P_{s\bar{s}}=0.0364^2\simeq0.1\%$. This is because
of that on the one hand the transition matrix element between
$s^{3}$ and $s^{4}\bar{s}$ configurations is smaller than the other
ones, and on the other hand energy of the $s^{4}\bar{s}$
configuration is much larger than the three-quark $\Omega$ state.

\begin{figure}[t]
\begin{center}
\includegraphics[scale=1.3]{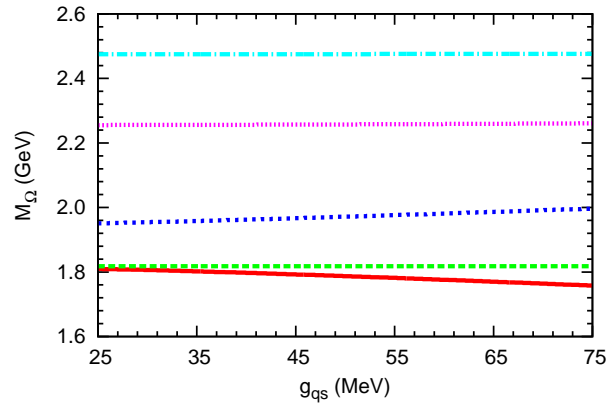}
\end{center}
\caption{\footnotesize (Color online) Energies of $\Omega$ resonances with spin $3/2$ as functions of
$g_{qs}$.
\label{g}}
\end{figure}

\begin{figure}[t]
\begin{center}
\includegraphics[scale=1.3]{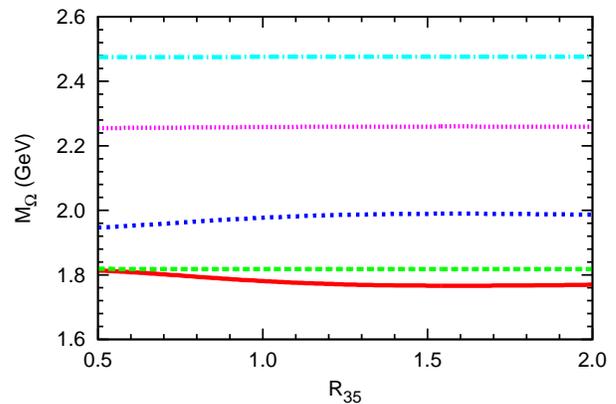}
\end{center}
\caption{\footnotesize (Color online) Energies of $\Omega$ resonances with spin $3/2$ as functions of
$R_{35}$.
\label{r35}}
\end{figure}

\subsection{Dependence of numerical results on parameters}
\label{free}

To show the dependence of mixing between three- and five-quark
$\Omega$ states with spin $3/2$, we present the energies of the
obtained states in Fig.~\ref{g} and Fig.~\ref{r35}, the former one
shows $M_{\Omega}$ as functions of the NJL interaction strength
$g_{qs}$ with $g_{qs}$ varying from $25$~MeV to $75$~MeV, and the
latter shows dependence of $M_{\Omega}$ on the ratio $R_{35}$ with
$R_{35}$ being in the range $0.5-2$. Note that here we just show the
dependence of mixing effects on parameters, since there is no mixing
between three- and five-quark configurations in the obtained
$\Omega$ states with spin $1/2$, so we only give the numerical
results for $\Omega$ states with spin $3/2$ in this section. In
addition, we keep the diagonal terms being constants when varying
the parameters.

As we can see in Fig.~\ref{g}, the energy of the lowest state shows
a little sensitivity on $g_{qs}$, $M_{\Omega}^{lowest}$ falls in the
range $1785\pm25$~MeV when $g_{qs}$ varying from $25$ to $75$ MeV,
this obtained energy is higher than the one
$M_{\Omega}^{lowest}=1750\pm50$~MeV in~\cite{An:2013zoa} by taking
the quark-antiquark pair creation mechanism to be the instanton
interaction induced one, since the transition matrix elements
between three- and five-quark $\Omega$ configurations in present
work are smaller than those obtained in~\cite{An:2013zoa}, as we
have discussed in Sec.~\ref{fixed}. While in any case, the obtained
value for $M_{\Omega}^{lowest}$ is decreased to be lower than energy
of the lowest spin $1/2$ state in present model, this conclusion is
consistent with that in the instanton-induced interaction model. On
the other hand, energies of the other states are not so sensitive to
$g_{qs}$, especially the highest two states, whose energies are even
constants, this is because of that the mixing effects are very small
in these two states, as shown in Table~\ref{num1}. In fact, energy
of the next-to-lowest state is also not sensitive to $g_{qs}$, as we
can see in Fig.~\ref{g}. This is because of that the main components
is this state are the first two five-quark configurations, whose
masses are very close to each other, lying at $\sim1820$, and mixing
between three- and five-quark configurations in this state is not
strong, as shown in Table~\ref{num1}, there is only $\sim0.1\%$
three-quark component in this state.

Fig.~\ref{r35} shows almost the same features as Fig.~\ref{g}, only energies of two
obtained states are sensitive to $R_{35}$, and the lowest energy falls in the range
$1790\pm20$~MeV.

Comparing to the numerical results in Ref.~\cite{An:2013zoa},  one can find that
the present obtained energies for spin $3/2$ states are very different to those
obtained by taking both quark-quark hyperfine interaction
and quark-antiquark pair creation to be the instanton-induced ones,
almost all the present obtained energies are lower than those in~\cite{An:2013zoa}.
This is because of that on the one hand the instanton-induced hyperfine interaction
is very different to the present NJL interaction induced one, the most obvious two
differences are those firstly the instanton-induced hyperfine interaction can only exist between
two quarks whose flavor wave function is antisymmetric, and secondly instanton-induced hyperfine
interaction leads to strong mixing between five-quark configurations with light and strange quark-antiquark
pairs. And on the other hand the transition matrix elements between three- and five-quark
spin $3/2$ states in instanton-induced quark-antiquark pair creation model
are larger than present ones.

While in any case, the mixing between three- and five-quark spin
$1/2$ $\Omega$ states is very small and negligible in INS model,
and is $0$ in NJL model,
while both INS and NJL induced quark-antiquark pair creation
mechanism result in strong mixing between three- and five-quark spin $3/2$
$\Omega$ states, and the mixing decreases energy of the lowest spin $3/2$
to lower than that of the lowest spin $1/2$ state.

\section{Conclusion}

\label{sec:end}

In this work, we investigate the mixing of the low-lying three- and
five-quark $\Omega$ configurations with negative parity in a NJL
interaction induced quark-antiquark pair creation model. Hyperfine
interaction between quarks is also taken to be based on the NJL
interaction for model consistency. Numerical results show that the
three- and five-quark configurations with spin $1/2$ do not mix to
each other in present model, this is because of that the spin
structure of the five-quark $\Omega$ states with spin $1/2$ results
in vanishing matrix elements for the transition $sss\leftrightarrow
sssq\bar{q}$. This is consistent with the results obtained in the
instanton induced quark-antiquark pair creation model, within which
mixing between three- and five-quark $\Omega$ states with
spin-parity $1/2^{-}$ is very small and
negligible~\cite{An:2013zoa}.

The NJL interaction induced quark-antiquark pair creation results in strong
mixing between three- and five-quark spin $3/2$ $\Omega$ states, and the mixing
decreases energy of the lowest state to $1785\pm25$~MeV, which is lower
than energy of the lowest spin $1/2$ state. On the other hand, mixing between
$s^{3}$ and $s^{4}\bar{s}$ configurations is very small.
This is also consistent with our previous
predictions within the instanton-induced quark-antiquark pair creation model.

While at present time, the experimental data of the $\Omega$
resonances spectrum is very poor, so we cannot say which model is
more appropriate. Recently, BESII Collaboration at Beijing Electron
Positron Collider (BEPC) reported an interesting result that
$\psi(2S)\to\Omega\bar{\Omega}$ was observed with a branch fraction
of $(5\pm 2)\times 10^{-5}$~\cite{Ablikim:2012qp}. Now with the
upgraded BEPC, {\sl i.e.}, BEPCII, BESIII
Collaboration~\cite{Asner:2008nq} is going to take billions of
$\psi(2S)$ events, which is two orders of magnitude higher than what
BESII experiment got. We hope the BESIII
experiment~\cite{Asner:2008nq} will provide us more information
through $\psi(2S)\to\bar\Omega\Omega^*$ reaction.

\section{Acknowledgements}
We thank Qun Wang for the suggestion to
consider the NJL interaction and useful
discussions.
This work is supported by the National Natural Science Foundation of China under Grant
Nos. 11205164, 11035006, 11121092, 11261130311 (CRC110 by DFG and NSFC), the Chinese Academy of
Sciences under Project No. KJCX2-EW-N01 and the Ministry of Science and Technology of
China (2009CB825200).



\end{document}